# Non-Markovian Effects on the Brownian Motion of a Free Particle

A. O. Bolivar

Instituto Mário Schönberg de Física-Matemática-Filosofia,

Ceilândia, Caixa Postal 7316, DF 72225-971, Brazil

E-mail: bolivar@cbpf.br

**Abstract** Non-Markovian effects upon the Brownian movement of a free particle in the presence as well as in the absence of inertial force are investigated within the framework of Fokker—Planck equations (Rayleigh and Smoluchowski equations). More specifically, it is predicted that non-Markovian features can enhance the values of the mean square displacement and momentum, thereby assuring the mathematical property of differentiability of the these physically observable quantities.

**Keywords**: Brownian Motion; Smoluchowski equation; Ornstein—Uhlenbeck process; Fokker—Planck equation; non-Markov effects

# 1 Introduction

The Brownian motion of a particle with mass $m$ and position $X = X(t)$ may be described by the Langevin equation [1]

$$m\,\frac{d^2X}{dt^2} = -\frac{dV(X)}{dX} - \gamma\,\frac{dX}{dt} + L(t), \tag{1}$$

where the inertial force $md^2X/dt^2$ offsets a set of three sorts of forces: a conservative force derived from an external potential $V(X)$; a linearly velocity-dependent dissipative force, $-\gamma dX/dt$, accounts for stopping the particle's motion; and a stochastic force $L(t)$, dubbed Langevin's force, responsible for activating the particle's movement. The damping constant $\gamma$ exhibits dimensions of mass per time.

As far as a free particle is concerned, the Langevin equation (1) may be written in terms of the linear momentum $P = P(t) = mdX/dt$ as

$$\frac{dP}{dt} = -\gamma\,\frac{P}{m} + L(t). \tag{2}$$

On the other hand, in the absence of inertial force, i.e., $md^2X/dt^2 = 0$, the free Brownian particle is described by the stochastic differential equation

$$\frac{dX}{dt} = \frac{1}{\gamma}L(t). \tag{3}$$

In the Gaussian approximation it is assumed that the Langevin force $L(t)$ present in (1—3) has the following statistical properties [2]

$$\langle L(t)\rangle = 0, \tag{4}$$

$$\langle L(t')L(t'')\rangle = D\delta(t' - t''), \tag{5}$$

where $D$ is a constant and the average value $\langle \dots \rangle$ is evaluated from the probability distribution function associated with the environmental random function $L(t)$. Equation (4) characterizes the irregularity feature of the Langevin force and Eq. (5) means that the interaction between the tagged particle and a generic environment is deemed to be Markovian in the sense that the autocorrelation function (5) is delta-correlated (white noise).

For thermal open systems characterized by the temperature $T$ and the Boltzmann constant $k_B$, the Ornstein-Uhlenbeck process (2), along with (4) and (5), leads to the mean square momentum [3]

$$\Delta P(t) = \sqrt{mk_BT\left(1 - e^{-\frac{2\gamma}{m}t}\right)}, \tag{6}$$

whereas the Fokker—Planck equation (the so-called diffusion equation) generated by the stochastic differential equation (3) with the statistical properties (4) and (5) yields the Einstein's mean square displacement [4]

$$\Delta X(t) = \sqrt{\frac{2k_B T}{\gamma} t}\,. \tag{7}$$

Both Markovian results (6) and (7) are non-differentiable functions at $t = 0$. Hence, it has been claimed that physically there is no force applied to a free Brownian particle in the presence of inertial forces, as well as no velocity in the absence of inertial forces [5,6]. This fact suggests that the concept of trajectory of a free Brownian particle seems to be an elusive feature in the Markovian regime.

The purpose of the present paper is to show how non-Markovian effects upon a free Brownian particle are responsible for the differentiability property of the mean square displacement (6) and momentum (7), thereby predicting an increase of the values of these physically observable quantities.

Our paper discusses the Brownian motion of a free particle reckoning with two physical situations described by Fokker—Planck equations: In Sect. 2 the case of a Brownian particle in the presence of inertial forces (Ornstein—Uhlenbeck stochastic process) described by the non-Markovian Rayleigh equation is examined, whereas in Sect. 3 we solve our non-Markovian Smoluchowski equation describing a free Brownian particle in the absence of inertial effects. Concluding remarks are presented in Sect. 4. In addition, two appendices are included.

## 2 Brownian Motion in the Presence of Inertial Force

A particle with mass $m$, position $X = X(t)$ immersed in a generic environment (e.g., a non-thermal fluid) undergoes a jittering movement dubbed Brownian motion. The motion of this tagged particle may be mathematically described by the stochastic differential equation (1) (the so-called Langevin equation) in the form

$$m\frac{d^2X}{dt^2} = -\frac{dV(X)}{dX} - \gamma\frac{dX}{dt} + \gamma b\Psi(t), \tag{8}$$

with the Langevin force given by $L(t) = \gamma b\Psi(t)$, where the parameter $b$ controls the environmental influence (fluctuations) upon the Brownian particle. Because the term $b\Psi(t)$ in (8) has dimensions of velocity, i.e., $[length \times time^{-1}]$, we can readily check that $b$ may be expressed in dimensions of $[length \times time^{-1/2}]$ and the function $\Psi(t)$ in dimensions of $[time^{-1/2}]$.

From the mathematical viewpoint the quantities $X = X(t)$ and $\Psi = \Psi(t)$ in the Langevin equation (8) are interpreted as random variables belonging to the Kolmogorov probability space [7] in the sense that there exists a probability distribution function, $\mathcal{F}_{X\Psi}(x,\psi,t)$, associated with the stochastic system $\{X,\Psi\}$, expressed in terms of the possible values $x = \{x_i\}$ and $\psi = \{\psi_i\}$, with $i \geq 1$, distributed about the sharp values $q$ and $\varphi$ of $X$ and $\Psi$, respectively. In addition, the average values of $X$ and $\Psi$ are expressed as

$$\langle X\rangle = \iint_{-\infty}^{+\infty} x\mathcal{F}_{X\Psi}(x,\psi,t)dxd\psi, \tag{9}$$

$$\langle \Psi\rangle = \iint_{-\infty}^{+\infty} \psi\mathcal{F}_{X\Psi}(x,\psi,t)dxd\psi. \tag{10}$$

For a free Brownian particle the Langevin equation (8) may be expressed in terms of the concept of linear momentum $P = P(t) = mdX/dt$ as

$$\frac{dP}{dt} = -\gamma\frac{P}{m} + \gamma b\Psi(t). \tag{11}$$

Noticing that

$$\lim_{\varepsilon\to 0}\int_t^{t+\varepsilon} \langle \Psi(t')\rangle dt' = 0, \tag{12}$$

with

$$\langle\Psi(t)\rangle = \int_{-\infty}^{+\infty} \psi \mathcal{F}_{\Psi}(\psi, t) d\psi,$$

the Fokker—Planck equation generated by (11) in the Gaussian approximation reads (see Appendix A)

$$\frac{\partial g(p,t)}{\partial t} = \gamma \frac{\partial}{\partial p}\left\{\left[\frac{p}{m} - b\langle\Psi(t)\rangle\right] g(p,t)\right\} + \mathcal{D}_p(t)\frac{\partial^2 g(p,t)}{\partial p^2} \quad (13)$$

in terms of the probability distribution function $g(p,t)$, the mean value of $\Psi(t)$

$$\langle\Psi(t)\rangle = \lim_{\varepsilon\to 0}\frac{1}{\varepsilon}\int_{t}^{t+\varepsilon} \langle\Psi(t')\rangle dt', \quad (14)$$

and the diffusion coefficient

$$\mathcal{D}_p(t) = \gamma\mathcal{E}(t) \quad (15)$$

where the function $\mathcal{E}(t)$, given by

$$\mathcal{E}(t) = \frac{\gamma b^2}{2} I(t) = \frac{\gamma b^2}{2}\lim_{\varepsilon\to 0}\frac{1}{\varepsilon}\iint_{t}^{t+\varepsilon} \langle\Psi(t')\Psi(t'')\rangle dt' dt'', \quad (16)$$

has dimensions of energy, i.e., $[mass \times length^2 \times time^{-2}]$. Hence we call $\mathcal{E}(t)$ the diffusion energy responsible for the Brownian motion of the particle dipped in a generic environment.

An outstanding feature underlying the diffusion energy concept (16), which fulfils the validity condition

$$0 < \mathcal{E}(t) < \infty, \quad (17)$$

is that it conveys in tandem fluctuation and dissipation phenomena through the function $I(t)$ and the friction constant $\gamma$, respectively. We can then state that (16) sets up a general fluctuation—dissipation relationship underlying all open systems described by the Langevin equation (11) and its corresponding Fokker—Planck equation (13). Both cases $\mathcal{E}(t) = 0$ and $\mathcal{E}(t) = \infty$ are not concerned, for they may violate the validity condition of the fluctuation—dissipation relation (17). The former case may lead to dissipation without fluctuation, whilst the latter one may give rise to fluctuation without dissipation.

From the physical point of view the pivotal issue inherent in theory of Brownian motion is to determine the transport coefficient, that is, the time-dependent diffusion coefficient (15). At $t = 0$ the answer to this question seems to be fairly straightforward, for the diffusion energy is null, i.e., $\mathcal{E}(0) = \gamma\mathcal{D}(0) = 0$, meaning that there is no diffusive motion, $\mathcal{D}(0) = 0$, associated with the initial condition $g(p, t = 0) = \delta(p)$ to our Fokker—Planck equation (13). On the other hand, at the long-time regime, $t \to \infty$, we assume that the steady diffusion energy can be identified with the thermal energy, $k_B T$, at thermodynamic equilibrium, i.e.,

$$\lim_{t\to\infty} \mathcal{E}(t) = \frac{\gamma b^2}{2} = k_B T, \tag{18}$$

with the dimensionless function $I(t)$ in (15) having the asymptotic behavior

$$\lim_{t\to\infty} I(t) = 1. \tag{19}$$

The physical significance of condition (19) has to do with the fact that environmental fluctuations decay to Markovian correlations, i.e., $I(t)$ displays a local (short) range behavior in the steady regime. By contrast, non-Markovian effects show up at the range $0 < t < \infty$. The correlational function $I(t)$in (16) fulfilling condition (19) can be built up as (see Appendix B)

$$I(t) = 1 - e^{-\frac{t}{t_c}}, \tag{20}$$

where the correlation time $t_c$ accounts for non-Markovian effects upon the Brownian particle.

It is worth underscoring that the identity (18) expresses the principle of energy conservation, for the diffusion energy $\mathcal{E}(\infty) = \gamma b^2/2$ comes from the Markovian Brownian dynamics (Langevin and Fokker—Planck equations at the steady regime), whereas the thermal energy $k_B T$ is a physical quantity stemming from the environment at thermodynamic equilibrium (equation of state for perfect gases: $\mathcal{P}\mathcal{V} = \mathcal{N} k_B T$, where $\mathcal{N}$ is the number of particles within the volume $\mathcal{V}$ under the pressure $\mathcal{P}$ and temperature $T$; the constant $k_B$ denotes the Boltzmann constant and displays dimensions of energy per temperature).

As a consequence of the conservation of energy (18), from (15) at the thermal equilibrium we can readily derive the Ornstein—Uhlenbeck diffusion constant as

$$\mathcal{D}_p(\infty) = \gamma k_B T, \tag{21}$$

and the parameter $b$ turns out to be of thermal nature, i.e.,

$$b = \pm\sqrt{\frac{2k_B T}{\gamma}}. \tag{22}$$

Using (22), (20), (16), and (15) our Fokker—Planck equation (13) reads

$$\frac{\partial g(p,t)}{\partial t}=\gamma\frac{\partial}{\partial p}\left\{\left[\frac{p}{m}\mp\sqrt{\frac{2k_BT}{\gamma}}\langle\Psi(t)\rangle\right]g(p,t)\right\}+\gamma k_BT\left(1-e^{-\frac{t}{t_c}}\right)\frac{\partial^2 g(p,t)}{\partial p^2}. \quad (23)$$

or

$$\frac{\partial g(p',t)}{\partial t}=\frac{\gamma}{m}\frac{\partial}{\partial p'}\{p'g(p',t)\}+\gamma k_BT\left(1-e^{-\frac{t}{t_c}}\right)\frac{\partial^2 g(p',t)}{\partial p'^2}, \quad (24)$$

in terms of the thermal position

$$p^{'}=p\mp m\sqrt{\frac{2k_BT}{\gamma}}\langle\Psi(t)\rangle. \quad (25)$$

According to (23) a particle immersed in a heat bath undergoes a Brownian motion owing to the diffusion energy stemming from the thermal reservoir, $k_BT$, multiplied by the correlational effects present in the function $I(t)=\left(1-e^{-t/t_c}\right)$. In general, the non-equilibrium thermal energy $\mathcal{E}(t)=k_BT\left(1-e^{-t/t_c}\right)$ is different from the equilibrium thermal energy, $k_BT$. They equal only at the steady regime (18). In this thermal equilibrium state the stochastic process is called normal diffusion, whereas in the non-equilibrium thermal regime characterized by $0<I(t)<1$ the Brownian motion is said to be subdiffusive.

It is worth pointing out that the diffusion constant (21) has been derived without specifying *ab initio* the form of the autocorrelation function of the Langevin force,

$$\langle L(t')L(t'')\rangle=2\gamma k_BT\langle\Psi(t')\Psi(t'')\rangle=2\mathcal{D}_p(\infty)\langle\Psi(t')\Psi(t'')\rangle,$$

as well as its average value, i.e.,

$$\langle L(t)\rangle=\pm\sqrt{2\gamma k_BT}\langle\Psi(t)\rangle=\pm\sqrt{2\mathcal{D}_p(\infty)}\langle\Psi(t)\rangle.$$

The crucial point is the long time behavior (19). This means that the arbitrariness as to the form of the statistical properties of the Langevin force $L(t)$ leaves room to incorporate non-Markovian and averaging effects into the study of Brownian motion through the functions $I(t)$ and $\langle\Psi(t)\rangle$ in the Fokker—Planck equation (23). So, starting from the initial condition $g(p^{'},t=0)=\delta(p')$ a non-equilibrium solution to (23) reads

$$g(p',t)=\frac{1}{\sqrt{4\pi\mathcal{G}(t)}}e^{-\frac{p'^2}{4\mathcal{G}(t)}}, \quad (26)$$

where

$$\mathcal{G}(t) = \frac{mk_BT}{2}\left[1 - e^{-\frac{2\gamma}{m}t} - \frac{2\gamma t_c}{(2\gamma t_c - m)} e^{-\frac{t}{t_c}}\right]. \tag{27}$$

The probability distribution function (26), which is expressed in terms of the evolution time $t$, the relaxation time $t_r = m/\gamma$, and the correlation time $t_c$, gives rise to

$$\langle P' \rangle = 0, \tag{28}$$

$$\langle P'^2 \rangle = 2\mathcal{G}(t) = mk_BT\left[1 - e^{-\frac{2\gamma}{m}t} - \frac{2\gamma t_c}{(2\gamma t_c - m)} e^{-\frac{t}{t_c}}\right]. \tag{29}$$

Making use of (29) the mean mechanical energy associated with the free Brownian particle is given by

$$\langle E(t) \rangle = \frac{\langle P'^2 \rangle}{2m} = \frac{k_BT}{2}\left[1 - e^{-\frac{2\gamma}{m}t} - \frac{2\gamma t_c}{(2\gamma t_c - m)} e^{-\frac{t}{t_c}}\right] \tag{30}$$

which reduces to

$$\langle E(t) \rangle = \frac{k_BT}{2}\left(1 - e^{-\frac{2\gamma}{m}t}\right) \tag{31}$$

in the Markovian limit.

The mean square momentum, $\Delta P'(t) = \sqrt{\langle P'^2 \rangle - \langle P' \rangle^2}$, reads

$$\Delta P'(t) = \Delta P(t) = \sqrt{mk_BT\left[1 - e^{-\frac{2\gamma}{m}t} - \frac{2\gamma t_c}{(2\gamma t_c - m)} e^{-\frac{t}{t_c}}\right]}, \tag{32}$$

meaning that the average value of $\Psi(t)$ in Eq. (25) has no influence upon the physically observable quantity (32). It is readily to check that the quantity

$$\left.\frac{d\Delta P(t)}{dt}\right|_{t=0} = \frac{2\gamma^2 k_BT t_c}{(2\gamma t_c - m)} \sqrt{\frac{m - 2\gamma t_c}{2mk_BT\gamma t_c}} \tag{33}$$

does not diverge provided that $0 < t_c < m/2\gamma$, thereby implying that non-Markovian effects are responsible for enhancing the mean square momentum (32). Yet in the Markovian limit we find from (32) the result

$$\Delta P(t) = \sqrt{mk_BT\left(1 - e^{-\frac{2\gamma}{m}t}\right)} \tag{34}$$

that is non-differentiable at $t = 0$. Non-Markovian effects in (32) therefore account for the existence of the concept of force acting upon a free Brownian particle in the presence of inertial force, since $0 < t_c < m/2\gamma$.

At the equilibrium regime, from (30) the energy equipartition is readily obtained as

$$\langle E(\infty)\rangle = \frac{k_B T}{2}, \tag{35}$$

while from the non-equilibrium solution (26) we derive the Maxwell—Boltzmann probability distribution function at thermal equilibrium

$$g(p') = \frac{1}{\sqrt{2\pi m k_B T}} e^{-\frac{p'^2}{2mk_B T}} \tag{36}$$

for the thermal momentum (25).

# 3 Brownian Motion in the Absence of Inertial Force

Now we start from the Langevin equation (8) in the Smoluchowski limit (the large friction case)

$$\left|-\gamma\frac{P}{m} + \gamma b\Psi(t)\right| \gg \left|\frac{dP}{dt}\right|, \tag{37}$$

such that inertial effects in (8) can be negligible, i.e., $md^2X/dt^2 = 0$. So the Brownian motion is approximated by the stochastic differential equation

$$\frac{dX}{dt} = -\frac{1}{\gamma}\frac{dV(X)}{dX} + b\Psi(t), \tag{38}$$

which gives rise to the following Fokker—Planck equation on configuration space in the Gaussian approximation (see Appendix A)

$$\frac{\partial f(x,t)}{\partial t} = -\frac{\partial}{\partial x}\left[-\frac{1}{\gamma}\frac{dV(x)}{dx} + b\langle\Psi(t)\rangle\right] f(x,t) + \mathcal{D}_x(t)\frac{\partial^2 f(x,t)}{\partial x^2}, \tag{39}$$

where the time-dependent diffusion coefficient is given by

$$\mathcal{D}_x(t) = \frac{\mathcal{E}(t)}{\gamma}, \tag{40}$$

with the diffusion energy $\mathcal{E}(t)$ being given by (16). Averaging effects of the random velocity, $b\langle\Psi(t)\rangle$, upon the drift coefficient bring about the time-dependent effective velocity

$$v_{\text{eff}}(x,t) = -\frac{1}{\gamma}\frac{dV(x)}{dx} + b\langle\Psi(t)\rangle. \tag{41}$$

For thermal systems at thermodynamic equilibrium in which (18) is valid, the diffusion constant (40) becomes the Einstein relation

$$\mathcal{D}_x(\infty) = \frac{k_B T}{\gamma}. \tag{42}$$

Accordingly, the non-Markovian Fokker—Planck equation (39) in the presence of thermal fluctuations reads

$$\frac{\partial f(x,t)}{\partial t} = \frac{\partial}{\partial x}\left[\frac{1}{\gamma}\frac{dV(x)}{dx} \mp \sqrt{\frac{2k_B T}{\gamma}}\langle\Psi(t)\rangle\right] f(x,t) + \frac{k_B T}{\gamma} I(t) \frac{\partial^2 f(x,t)}{\partial x^2}. \tag{43}$$

For $\langle\Psi(t)\rangle = 0$ and $I(t) = 1$, the Fokker—Planck equation (43) yields the equation of motion

$$\frac{\partial f(x,t)}{\partial t} = \frac{1}{\gamma}\frac{\partial}{\partial x}\left[\frac{dV(x)}{dx} f(x,t)\right] + \frac{k_B T}{\gamma}\frac{\partial^2 f(x,t)}{\partial x^2}, \tag{44}$$

early found out by Smoluchowski [8] and widely investigated in the literature [9—14]. The Smoluchowski equation (44) is defined for Markovian correlations and no averaging effects in contrast to our Fokker—Planck equation (43) which takes into account non-Markovian effects in view of the time-dependent function $I(t)$ as well as averaging effects present in the drift coefficient.

Again, our account of the Einstein—Langevin stochastic approach presented above has pointed out that the main upshot of Einstein—the Founding Father of Brownian motion theory—rests on the fact that his diffusion coefficient (42) is a consequence of the conservation of energy (43) for thermal open systems at thermodynamic equilibrium.

It is worth recalling that the Einstein relation (42) has been attained without specifying *ab initio* the form of the autocorrelation function of the Langevin force,

$$\langle L(t)L(t')\rangle = 2\gamma^2 \frac{k_B T}{\gamma}\langle\Psi(t')\Psi(t'')\rangle = 2\gamma^2 \mathcal{D}_x(\infty)\langle\Psi(t')\Psi(t'')\rangle,$$

as well as its average value, i.e.,

$$\langle L(t)\rangle = \pm\gamma\sqrt{2\frac{k_B T}{\gamma}}\langle\Psi(t)\rangle = \pm\gamma\sqrt{2\mathcal{D}_x(\infty)}\langle\Psi(t)\rangle.$$

3.1 Non-Markovian Brownian Motion of a Free Particle

As far as the non-inertial Brownian motion of a free particle is concerned, our non-Markovian Smoluchowski equation (43) with (20) at point $x^{'}$ reads

$$\frac{\partial f(x',t)}{\partial t} = \mp\sqrt{\frac{2k_BT}{\gamma}}\langle\Psi(t)\rangle\frac{\partial f(x',t)}{\partial x'} + \frac{k_BT}{\gamma}\left(1 - e^{-\frac{t}{t_c}}\right)\frac{\partial^2 f(x',t)}{\partial x'^2}. \qquad (45)$$

Starting from the deterministic initial condition

$$f(x',t=0) = \delta(x'), \qquad (46)$$

characterized by $\mathcal{E}(0) = 0$ and $\langle\Psi(0)\rangle = 0$, we obtain the following time-solution to (45)

$$f(x,t) = \sqrt{\frac{\gamma}{4\pi A(t)}}e^{-\frac{\gamma x^2}{4A(t)}}, \qquad (47)$$

in terms of the thermal position

$$x = x^{'} \mp \sqrt{\frac{2k_BT}{\gamma}}\int\langle\Psi(t)\rangle dt, \qquad (48)$$

and the function

$$A(t) = k_BT\int I(t)dt = k_BT\left(t + t_c e^{-\frac{t}{t_c}}\right). \qquad (49)$$

Besides reducing to (46) at $t = 0$ (along with $t_c = 0$), solution (47) yields

$$\langle X\rangle = 0, \qquad (50)$$

and

$$\langle X^2\rangle = \frac{2k_BT}{\gamma}\left(t + t_c e^{-\frac{t}{t_c}}\right), \qquad (51)$$

where the stochastic position $X = X(t)$ corresponding to (48) is given by

$$X = X^{'}(t) \mp \sqrt{\frac{2k_BT}{\gamma}}\int\langle\Psi(t)\rangle dt.$$

Accordingly, the mean square displacement, $\Delta X(t) = \sqrt{\langle X^2\rangle - \langle X\rangle^2}$, reads

$$\Delta X(t) = \Delta X'(t) = \sqrt{\frac{2k_BT}{\gamma}\left(t + t_c e^{-\frac{t}{t_c}}\right)}. \qquad (52)$$

We notice that averaging effects are unobservable, for they do have no influence upon the physically measurable quantity (52), albeit they can bring about a shift in the position (48). By contrast, non-Markovian effects account for enhancing the mean square displacement (52) of a free Brownian particle.

For very short times $t \ll t_c$, quantity (52) reduces to a constant value given by

$$\Delta X = \sqrt{\frac{2k_B T}{\gamma} t_c} \tag{53}$$

which is differentiable, i.e., $d\Delta X/dt = 0$. On the other hand, in the Markovian limit, $t_c \to 0$, quantity (52) reduces to the Einstein's renowned upshot

$$\Delta X(t) = \sqrt{\frac{2k_B T}{\gamma} t}. \tag{54}$$

that is non-differentiable at $t = 0$, i.e.,

$$\left.\frac{d\Delta X(t)}{dt}\right|_{t=0} = \left.\sqrt{\frac{k_B T}{2\gamma t}}\right|_{t=0} \to \infty. \tag{55}$$

Hence, it is claimed that there is no concept of velocity of a Brownian particle in the strong friction regime [5,6]. In fact, the physically embarrassing outcome (55) is a consequence of neglecting non-Markovian features in (52).

## 4 Concluding Remarks

In this paper we have examined the Brownian motion of a free particle immersed in a thermal reservoir reckoning with non-Markovian effects. In Sec. 2 we have shown that the mean square momentum (32) is a differentiable function in the presence of inertial force provided that non-Markovian effects increase the value of (32).

In the absence of inertial force in Sec. 3 it has been predicted that non-Markovian effects enhance the mean square displacement (52), thereby assuring the mathematical property of differentiability of this physically observable quantity.

It is worth point out that our chief results have been obtained without making use of the generalized Langevin equation

$$m\,\frac{d^2X}{dt^2} = -\frac{dV(X)}{dX} - \int_0^t \beta(t' - t'')\,\frac{dX(t')}{dt}\,dt' + L(t)$$

put forward by Mori [15] on the basis of a close relationship between memory effects ingrained in the friction kernel $\beta(t' - t'')$ and non-Markovian effects (colored noise) showing up in the autocorrelation function $\langle L(t')L(t'')\rangle$, given by

$$\langle L(t')L(t'')\rangle = \frac{D}{\gamma}\beta(t' - t'').$$

By contrast, our approach has predicted that *non-Markovian effects independent of memory effects* can be physically measured for a free Brownian particle immersed in a heat bath by means of the mean square displacement in the absence of inertial force as well as the mean square momentum in the presence of the inertial force. Surprisingly, this feature standing out in the Einstein—Langevin framework has been overlooked in the centenary literature about Brownian motion [1—20].

## Appendix A: Generalized Fokker—Planck Equations

Let us consider the stochastic differential equation

$$\frac{dZ}{dt} = -\frac{1}{\gamma}\frac{d\mathcal{V}(Z)}{dZ} + a\Psi(t). \tag{A1}$$

For $Z = P$, $\mathcal{V}(P) = \gamma^2 P^2/2m$, and $a = \gamma b$ Eq. (A1) reduces to the Langevin stochastic equation (11), whereas for $Z = X$, $\mathcal{V} = V(X)$, and $a = b$ we obtain the Langevin equation (44).

Equation (A1) gives rise to the Kolmogorov equation [7,11,17,18]

$$\frac{\partial f(z,t)}{\partial t} = \mathbb{K} f(z,t), \tag{A2}$$

where the Kolmogorovian operator $\mathbb{K}$ acts upon the probability distribution function $f(z,t)$ according to

$$\mathbb{K} f(z,t) = \sum_{k=1}^{\infty} \frac{(-1)^k}{k!}\frac{\partial^k}{\partial z^k}[A_k(z,t) f(z,t)], \tag{A3}$$

the coefficients $A_k(z,t)$ being given by

$$A_k(z,t) = \lim_{\varepsilon\to 0}\left[\frac{\langle(\Delta Z)^k\rangle}{\varepsilon}\right], \tag{A4}$$

where the average values, $\langle(\Delta Z)^k\rangle$, are to be calculated about the sharp values z' from the probability distribution function

$$\mathcal{F}_{Z\Psi}(z,\psi,t) = \delta(z - z')\mathcal{F}_{\Psi}(\psi,t). \tag{A5}$$

According to the Pawula theorem [18], if the coefficients $A_k(z,t)$ in (A4) are finite for every $k$ and if $A_k(z,t) = 0$ for some even $k$, then $A_k(z,t) = 0$ for all $k \geq 3$, thereby assuring the positivity of the probability density function $f(z,t)$. So, if $(z_2 - z_1)^3 \sim 0$ such that $(z_2 - z_1)^4 = 0$, it follows then that $A_k(z,t) = 0$, $k \geq 3$, where $z_2 = z(t+\varepsilon)$ and $z_1 = z(t)$. Accordingly, the Kolmogorov equation (A2) can be approximated by the Fokker—Planck equation

$$\frac{\partial f(z,t)}{\partial t} = -\frac{\partial}{\partial z}[A_1(z,t) f(z,t)] + \frac{1}{2}\frac{\partial^2}{\partial z^2}[A_2(z,t) f(z,t)], \tag{A6}$$

where the drift coefficient is given by

$$A_1(z,t) = \lim_{\varepsilon\to 0}\left[\frac{\langle\Delta Z\rangle}{\varepsilon}\right] = -\frac{1}{\gamma}\frac{d\mathcal{V}}{dz} + a\langle\Psi(t)\rangle \tag{A7}$$

and the diffusion coefficient reads

$$A_2(z,t) = \lim_{\varepsilon\to 0}\left[\frac{\langle(\Delta Z)^2\rangle}{\varepsilon}\right] = a^2 \lim_{\varepsilon\to 0}\frac{1}{\varepsilon}\iint_t^{t+\varepsilon} \langle\Psi(t')\Psi(t'')\rangle dt'dt'', \quad \text{(A8)}$$

with

$$\lim_{\varepsilon\to 0} \frac{1}{\varepsilon}\int_t^{t+\varepsilon} \langle\Psi(t')\rangle dt' = \langle\Psi(t)\rangle \quad \text{(A9)}$$

and

$$\lim_{\varepsilon\to 0} \int_t^{t+\varepsilon} \langle\Psi(t')\rangle dt' = 0. \quad \text{(A10)}$$

To evaluate the coefficients (A7) and (A8) we have used (A1) in the form

$$\Delta Z = Z(t+\varepsilon) - Z(t) = -\frac{\varepsilon}{\gamma}\frac{d\mathcal{V}}{dZ} + a\int_t^{t+\varepsilon} \Psi(t')dt'. \quad \text{(A11)}$$

For z= $p$, $\mathcal{V}(p) = \gamma^2 p^2/2m$, and $a = \gamma b$ the Fokker—Planck equation (A6) reduces to the Rayleigh equation (13), whereas for z= $x$, $\mathcal{V} = V(X)$, and $a = b$ (A6) leads to the Smoluchowski equation (45).

## Appendix B: The Time-dependent Diffusion Energy

The time-dependent diffusion energy

$$\mathcal{E}(t) = \mathcal{E}(\infty)I(t) \tag{B1}$$

in the Fokker—Planck equations (13) and (45) presents the correlational function

$$I(t) = \lim_{\varepsilon\to 0}\frac{1}{\varepsilon}\iint_{t}^{t+\varepsilon} \langle \Psi(t')\Psi(t'')\rangle dt'dt'' \,. \tag{B2}$$

On the condition that the autocorrelation function, $\langle \Psi(t')\Psi(t'')\rangle$, can be given by

$$\langle \Psi(t')\Psi(t'')\rangle = \left(1 - e^{-\frac{(t'+t'')}{2t_c}}\right)\delta(t'-t''), \tag{B3}$$

where $t_c$ is the correlation time of $\Psi(t)$ at times $t'$ and $t''$, it follows that (B2) becomes

$$I(t) = 1 - e^{-\frac{t}{t_c}}, \tag{B4}$$

reducing to $I(\infty) = 1$ in the steady regime.

The autocorrelation function (B3) defines a colored noise which in the Markovian limit, $t_c \to 0$, changes into the so-called white noise

$$\langle \Psi(t')\Psi(t'')\rangle = \delta(t'-t''), \tag{B6}$$

meaning that the stochastic function $\Psi(t)$ is delta-correlated [16,19]. It has been argued that the Markov property (B6) is a highly idealized feature [19], because the physical interaction between the Brownian particle and the thermal bath always comes about for a finite correlation time. By the same token, a decade ago van Kampen has laconically stated: "*Non-Markov is the rule, Markov is the exception*" [20]. Hence, our auto-correlation function (B3) seems to be a more realistic feature of the Langevin force.